\documentclass[pre,amsmath,amssymb,preprint,superscriptaddress]{revtex4}

\usepackage{amsmath}    
\usepackage{amsfonts}
\usepackage{amssymb}
\usepackage{graphicx}   
\usepackage{color}
\usepackage{dcolumn}
\usepackage{bm}

\begin{document}

\title{Quantum electrodynamic effects on counter-streaming instabilities in the whole \textbf{k} space}

\author{Antoine Bret}
\affiliation{ETSI Industriales, Universidad de Castilla-La Mancha, 13071 Ciudad Real, Spain}
 \affiliation{Instituto de Investigaciones Energ\'{e}ticas y Aplicaciones Industriales, Campus Universitario de Ciudad Real,  13071 Ciudad Real, Spain.}
 \email{antoineclaude.bret@uclm.es.}

\date{\today }

\begin{abstract}
In a recent work [Bret, EPL \textbf{135} (2021) 35001], quantum electrodynamic (QED) effects were evaluated for the two-stream instability. It pertains to the growth of perturbations with a wave vector oriented along the flow in a collisionless counter-streaming system. Here, the analysis is extended to every possible orientation of the wave vector. The previous result for the two-stream instability is recovered, and it is proved that even within the framework of a 3D analysis, this instability remains fundamentally 1D even when accounting for QED effects. The filamentation instability, found for wave vectors normal to the flow, is weakly affected by QED corrections. As in the classical case, its growth rate saturates at large $k_\perp$. The saturation value is found independent of QED corrections. Also, the smallest unstable $k_\perp$ is independent of QED corrections. Surprisingly, unstable modes found for oblique wave vectors do \emph{not} follow the same pattern. For some, QED corrections do reduce the growth rate. But for others, the same corrections increase the growth rate instead. The possibility for QED effects to play a role in un-magnetized systems is evaluated. Pair production resulting from gamma emission by particles oscillating in the exponentially growing fields, is not accounting for.
\end{abstract}

\maketitle

\section{Introduction}
Counter-streaming instabilities have been a central topic in plasma physics for nearly one century \cite{Langmuir1925,pierce1948}. Extreme plasma physics, on the other hand, has been rising during the last years \cite{Uzdensky2014,SilvaEPS2021}. It has to do with quantum electrodynamic (QED) effects that could appear in plasmas immersed in extreme electromagnetic fields. This new field of research is spurred by the advent of high-power lasers with which such effects could be observed \cite{danson2015,king2016}. Also involved are high energy astrophysics settings, like magnetars or pulsars, where magnetic fields of the order of the critical Schwinger field $B_{cr}=m^2c^3/q\hbar=4.4 \times 10^{13}$ Gauss, or even greater, are present \cite{Lai2015,Qu2021}.

In neutron stars or magnetars, particle beams flow along the field lines from one foot of the lines to the other. It is yet unclear how they are stopped when hitting the surface. There, counter-streaming instabilities could play a role in the dissipation of the kinetic energy \cite{Beloborodov2007}. Such instabilities have also been invoked in the context of pulsar emissions \cite{Gedalin2002,AsseoPPCF2003,Melrose2017}.

With fields $B_0$ of the order of $10^{12}$ G for neutron stars, the ratio $B_0/B_{cr}$ reaches 0.02 so that corrections described here are necessary. As for magnetars, the ratio $B_0/B_{cr}$ reaches 2 to 22 so that the present corrections are but preliminaries for even greater ones since the present treatment assumes $B_0/B_{cr} \ll 1$.

In the context of long Gamma-Ray-Bursts, some models propose a protomagnetar as central engine \cite{Usov1992,Bucciantini2008,Metzger2011}. As the jet it produces makes its way through the remaining of the progenitor star, counter-streaming instabilities in highly magnetized environment should be excited, especially in the inner part of the jet. Here, the field of the protomagnetar would also render necessary QED corrections.

It seems therefore natural to investigate QED effects on counter-streaming instabilities as they could be triggered in various high field environments. Recently, QED effects were studied for the two-stream instability (TSI) in the presence of a guiding magnetic field $B_0$ \cite{BretEPL2021}. It was found that for $B_0 \ll B_{cr}$, the growth rate is scaled down by a factor $\sqrt{1+\xi}$ with,
\begin{equation}\label{eq:xi}
\xi = \frac{\alpha}{9\pi}\left( \frac{B_0}{B_{cr}} \right)^2,
\end{equation}
where $\alpha \sim 1/137$ is the fine structure constant.

Still, it has been known for long that the TSI is not the only instability triggered in collisionless counter-streaming systems \cite{Watson,Bludman,fainberg}. While the TSI amplifies perturbations $\propto \exp(i \mathbf{k}\cdot \mathbf{r} - i \omega t)$ with $\mathbf{k}$ parallel to the flow, perturbations with $\mathbf{k}$ normal and even oblique to the flow can also grow as the filamentation instability (FI) and the oblique instability respectively. Depending on the parameters of the problem, various instabilities can dominate the unstable spectrum, and their hierarchy has been worked out for several kinds of systems \cite{BretPoPHierarchie,BretPRL2008,bretApJ2009,BretPoPReview,BretPoP2016,BretPoP2017}.

\begin{figure}
\begin{center}
 \includegraphics[width=.45\textwidth]{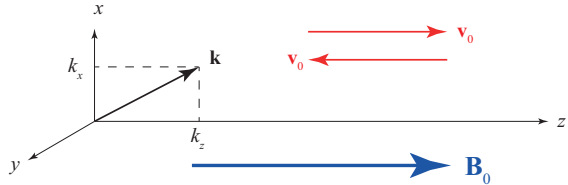}
\end{center}
\caption{System considered.}\label{fig:setup}
\end{figure}

The goal of this article is to extend the study of QED effects on the TSI to every possible perturbation.

We consider the simple system pictured on Figure \ref{fig:setup}. Two counter-streaming cold electron beams of identical density $n_0$ and opposed velocities $\pm \mathbf{v}_0$ stream along the $z$ axis. ``Cold'' here means that the thermal velocity spread $\Delta v$ in each beam satisfies $\Delta v \ll v_0$. A background of fixed ions ensures charge neutrality. An external magnetic field $\mathbf{B}_0$ is aligned with the flow. The system is therefore charge and current neutral at equilibrium, with no force acting on it since $\mathbf{v}_0 \times \mathbf{B}_0 = \mathbf{0}$.

Perturbations with wave vector $\mathbf{k}$ are applied. Since the system has cylindrical symmetry around $z$, we can choose the $x,y$ directions such that $k=(k_x,0,k_z)$ without loss of generality.

In Section \ref{sec:class} we briefly recall the structure of the classical calculation in order to clearly see where QED corrections come into play. Then in Section \ref{sec:QED}, QED corrections are worked out.

\section{Classical calculation}\label{sec:class}
Since the beams are assumed cold, a 2-fluids model can be implemented. We write the conservation equations for matter and momentum for the 2 beams,
\begin{eqnarray}
\frac{\partial n_i}{\partial t} + \nabla \cdot ( n_i \mathbf{v}_i) &=& 0, \label{eq:conser}  \\
\frac{\partial \mathbf{p}_i}{\partial t} + (\mathbf{v}_i\cdot \nabla) \mathbf{p}_i &=& q\left( \mathbf{E} + \frac{\mathbf{v}_i \times \mathbf{B}}{c}  \right), \label{eq:euler}
\end{eqnarray}
where the momentum $p_i$ reads $p_i=\gamma_i mv_i$ with $\gamma_i=(1-v_i^2/c^2)^{-1/2}$, $m$ the electron mass and $q$ its charge. We also write the Maxwell's equations involved in the calculation,
\begin{eqnarray}
\nabla \times \mathbf{E} &=& -\frac{1}{c}\frac{\partial \mathbf{B}}{\partial t}, \label{eq:Max1}\\
\nabla \times \mathbf{B} &=& \frac{1}{c}\frac{\partial \mathbf{E}}{\partial t} + \frac{4\pi}{c}\mathbf{J}. \label{eq:Max2}
\end{eqnarray}

The full calculation can be found in Refs. \cite{Godfrey1975,bretApJ2009}. It goes as follows. Assume first order perturbations $\propto \exp(i\mathbf{k}\cdot \mathbf{r} - i \omega t)$ of every quantity and linearize Eqs. (\ref{eq:conser}-\ref{eq:Max2}). Eqs. (\ref{eq:conser}-\ref{eq:euler}) give the 1st order density perturbation $n_{1i}$. Then Eq. (\ref{eq:euler}) gives the 1st order velocity perturbation $\mathbf{v}_{1i}$ in terms of $\mathbf{E}_1$, $\mathbf{B}_1$ and $\mathbf{B}_0$.  Finally, Eq. (\ref{eq:Max1}) gives $\mathbf{B}_1 = (c/\omega) \mathbf{k} \times \mathbf{E}_1$.

The classical current can then be computed as,
\begin{equation}\label{eq:Jclass}
\mathbf{J}_{class}(\mathbf{E}_1) = \underbrace{q\sum_i n_{0}\mathbf{v}_{0}}_{=0} + q\sum_i n_{0i}\mathbf{v}_{1i} + n_{1i}\mathbf{v}_{0i}.
\end{equation}
Finally, Eq. (\ref{eq:Max2}) is used to obtain,
\begin{equation}\label{eq:Maxwell}
 \mathbf{k} \times (\mathbf{k} \times \mathbf{E}_1) + \frac{\omega^2}{c^2}\left(\mathbf{E}_1 + \frac{4 i \pi}{\omega} \mathbf{J}_{class}(\mathbf{E}_1) \right)  = 0,
\end{equation}
that is, a tensorial equation of the form $\mathfrak{T}(\mathbf{E}_1)=\mathbf{0}$. The dispersion equation then comes through $\det\mathfrak{T}=0$. Note that when considering the TSI, the dispersion equation can be obtained from the scalar Poisson equation since the problem is 1D. Here, we need to use the vectorial Maxwell-Faraday equation for the current since the problem is 3D.

\begin{figure}
\begin{center}
 \includegraphics[width=.45\textwidth]{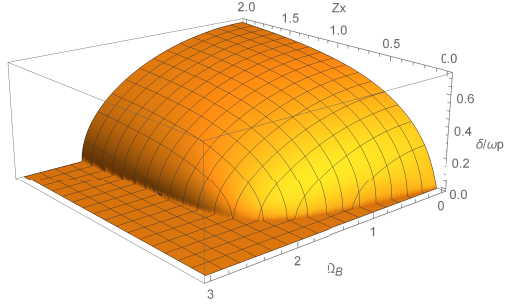}
\end{center}
\caption{Growth rate of the classical FI in terms of the field amplitude $\Omega_B$ and the wave vector $Z_x$. The growth rate saturates at large $Z_x$, and the instability is stabilized for $\Omega_B > \beta_0\sqrt{2\gamma_0}$.}\label{fig:filaclass}
\end{figure}

Although without major technical difficulties, calculations are lengthy. They can be performed with the \emph{Mathematica} Notebook described in Ref. \cite{BretCPC} in terms of the dimensionless variables,
\begin{eqnarray}\label{eq:var}
x &=& \frac{\omega}{\omega_p},~~\mathbf{Z} = \frac{\mathbf{k} v_0}{\omega_p},~~\beta_0 = \frac{v_0}{c},~~\gamma_0 = \frac{1}{\sqrt{1-\beta_0^2}}, \nonumber \\
\Omega_B &=& \frac{\omega_B}{\omega_p}, ~~\omega_B=\frac{|q|B_0}{mc},
\end{eqnarray}
with,
\begin{equation}
\omega_p^2 = \frac{4 \pi n_0 q^2}{m}.
\end{equation}

The \emph{Mathematica} Notebook used to compute the dispersion equation is provided as Supplemental  Material.

\begin{figure}
\begin{center}
 \includegraphics[width=.45\textwidth]{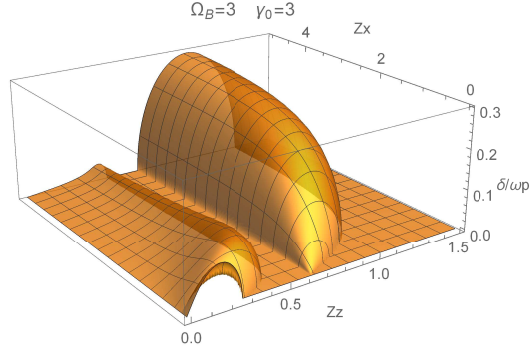}
\end{center}
\caption{Classical growth rate $\delta$ in terms of $Z_x$ and $Z_z$. The amplitude of the field $\Omega_B = 3 > \beta_0\sqrt{2\gamma_0} \sim 2.3$ is such that the filamentation instability is stabilized. The TSI is left unchanged with respect to the field-free case. The dominant unstable modes are now oblique.}\label{fig:class}
\end{figure}

The salient and known features of the classical case are,
\begin{itemize}
  \item For $Z_x=0$, that is $\mathbf{k}$ parallel to the flow, the TSI is left unchanged by the field since it has particles oscillating along the field, hence cancelling the Lorentz force.
  \item For $Z_z=0$, that is $\mathbf{k}$ normal to the flow, the growth rate $\delta$ of the FI is pictured on Figure \ref{fig:filaclass} in terms of $Z_x$ and $\Omega_B$.  For large $Z_x$, the growth rate saturates at \cite{Godfrey1975},
  \begin{equation}\label{eq:FilaClass}
  \delta(Z_x=\infty)=\frac{\sqrt{2 \beta_0^2 \gamma_0-\Omega_B^2}}{\gamma_0}.
  \end{equation}
   The most direct and interesting consequence of this equation is that FI is quenched beyond the critical value,
   \begin{equation}\label{eq:OmegaBc}
   \Omega_B > \beta_0\sqrt{2\gamma_0} \equiv \Omega_{Bc}.
   \end{equation}
  The field also stabilizes the small wavelengths fulfilling (see details in Section \ref{sec:FIqed}),
  \begin{equation}\label{eq:FilaStab}
  Z_x < \frac{\sqrt{2} }{\gamma_0^{3/2} }\frac{\beta_0 \Omega_B}{\sqrt{2 \beta^2 \gamma_0-\Omega_B^2}}.
  \end{equation}
  \item For both $Z_x, Z_z \neq 0$, Figure  \ref{fig:class} pictures a typical growth rate map in terms of $Z_x$ and $Z_z$. The amplitude of the field $\Omega_B = 3 > \beta_0\sqrt{2\gamma_0} \sim 2.3$ is such that the FI is stabilized. The TSI is left unchanged with respect to the field-free case. The dominant unstable modes are now oblique. We refer the reader to Ref. \cite{Godfrey1975} for the analysis of the unstable upper-hybrid-like modes with $Z_z \neq 0$ and $Z_x \gg 1$.
\end{itemize}

\section{QED calculation}\label{sec:QED}
QED corrections only modify the expression (\ref{eq:Jclass}) for the current \footnote{See for example Eqs. (1-8) of Ref. \cite{DiPiazzaPoP2007}.}. It now reads,
\begin{equation}\label{eq:JQED}
\mathbf{J} = \mathbf{J}_{class} + \mathbf{J}_{vac},
\end{equation}
where $\mathbf{J}_{class}$ is the classical result and,
\begin{equation}\label{eq:Jvac}
\mathbf{J}_{vac}= -\frac{1}{180\pi^2}\frac{\alpha}{c^2B_{cr}^2}\left(\nabla \times \mathbf{M} - \frac{\partial \mathbf{P}}{\partial t}  \right),
\end{equation}
where,
\begin{eqnarray}
  \mathbf{M}  &=& 2(E^2-c^2B^2)c\mathbf{B} - 7c (\mathbf{E}\cdot \mathbf{B})\mathbf{E}, \label{eq:M} \\
  \mathbf{P}  &=& 2(E^2-c^2B^2)\mathbf{E} + 7 c^2 (\mathbf{E}\cdot \mathbf{B})\mathbf{B}. \label{eq:P}
\end{eqnarray}
Since the $\nabla \times$ and the $\partial_t$ operators are linear, we can linearize $\mathbf{M}$ and $\mathbf{P}$ first, before applying these operators to obtain the linearized version of Eq. (\ref{eq:Jvac}). The linearization of Eqs. (\ref{eq:M},\ref{eq:P}) gives,
\begin{eqnarray}\label{eq:MPLin}
  \mathbf{M}  &=& -\left( \begin{array}{c}
                           0 \\
                           0 \\
                           2  B_0^3 c^3 +\frac{4  B_0^2 c^4 E_{1y} k_x}{\omega }
                         \end{array} \right) + \mathcal{O}(E_1^2),
   \\
  \mathbf{P}  &=&  \left( \begin{array}{r}
                           - 2  B_0^2 c^2 E_{1x}   \\
                           - 2  B_0^2 c^2 E_{1y}  \\
                             5 B_0^2 c^2 E_{1z}
                         \end{array} \right) + \mathcal{O}(E_1^2).
\end{eqnarray}
Since $E_{1x,y,z} \propto \exp(i\mathbf{k}\cdot \mathbf{r} - i \omega t)$ we find from Eq. (\ref{eq:Jvac}) the QED correction to the first order current,
\begin{equation}\label{eq:JvacLin}
\mathbf{J}_{1,vac} =  i \omega \frac{\xi}{20\pi}\left(\frac{ B_0}{B_{cr}}\right)^2 \left(  \begin{array}{r}
                           2  E_{1x}  \\
                           2 \left(1 - 2\frac{ c^2    k_x^2}{\omega^2 } \right)E_{1y}  \\
                           -5   E_{1z}
                         \end{array}  \right),
\end{equation}
where $\xi$ is the dimensionless parameter defined by Eq. (\ref{eq:xi}).

All the steps described for the classical case are therefore identical, except that we now need to add $\mathbf{J}_{1,vac}$ to the current equation (\ref{eq:Jclass}). The dispersion equation is eventually still of the form $\det\mathfrak{T}=0$ with now,
\begin{equation}\label{eq:T}
\mathfrak{T} = \mathfrak{T}_{Class} + \xi ~ \mathfrak{T}_{QED},
\end{equation}
with,
\begin{equation}\label{eq:Tqed}
\mathfrak{T}_{QED}=\left(
\begin{array}{ccc}
 -\frac{2}{5}   & 0 & 0 \\
 0 & \frac{2}{5}  \left(\frac{2 Z_x^2}{x^2 \beta_0^2}-1\right) & 0 \\
 0 & 0 & 1   \\
\end{array}
\right),
\end{equation}
so that the correction for all $\mathbf{k}$'s is of order $\xi$ and diagonal only.

Note that for all practical purposes, $\xi$ is an extremely small parameter since $B_0 \ll B_{cr}$ is required to write Eqs. (\ref{eq:JQED},\ref{eq:Jvac}) on the one hand (see \cite{DiPiazzaRMP2012} or \cite{weinberg2005quantum}, p. 32), while $\alpha \sim 1/137$ on the other hand. For example, with $B_0/B_{cr}=0.1$, Eq. (\ref{eq:xi}) gives $\xi = 2.5 \times 10^{-6}$. Even $\xi=0.1$ requires $B_0=19B_{cr}$, which is far out of range of the present theory.

Computing the dispersion equation is performed using a \emph{Mathematica} Notebook very similar to the one used for the classical case. The only difference is that the QED-corrected first order current (\ref{eq:JvacLin}) is added to equation (5) of Ref. \cite{BretCPC}.

Since the dispersion equation reads $\det \mathfrak{T} = 0$ where $\mathfrak{T}$ is now given by Eq. (\ref{eq:T}), an alternative to the forthcoming calculations would be to write,
\begin{equation}
\mathfrak{T}_{Class} + \xi ~\mathfrak{T}_{QED} = \mathfrak{T}_{Class} (I + \xi ~ \mathfrak{T}_{Class}^{-1}\mathfrak{T}_{QED}),
\end{equation}
so that,
\begin{eqnarray}
\det \mathfrak{T} &=& \det \mathfrak{T}_{Class} \times \det (I + \xi ~ \mathfrak{T}_{Class}^{-1}\mathfrak{T}_{QED}) \\
&=& \det \mathfrak{T}_{Class} \times \left\{1+\xi ~\mathrm{tr} (\mathfrak{T}_{Class}^{-1}\mathfrak{T}_{QED})   + O(\xi^2)\right\}, \nonumber
\end{eqnarray}
where $\mathrm{tr} ~\mathfrak{M}$ is the trace of the matrix $\mathfrak{M}$ and Jacoby's formula has been used to expand the determinant $\det (I + \xi ~ \mathfrak{T}_{Class}^{-1}\mathfrak{T}_{QED})$. The dispersion equation would then read at first order in $\xi$,
\begin{equation}
\det \mathfrak{T} = 0 ~~ \Rightarrow 1+\xi ~\mathrm{tr} (\mathfrak{T}_{Class}^{-1}\mathfrak{T}_{QED})  = 0.
\end{equation}
However, since the dispersion equation is to be solved for $x$ and not $\xi$, computations are not simpler than the ones explained from now.

We now review the consequences of the QED corrections on the various instabilities involved in the system.

\subsection{Two-Stream instability (TSI)}
Since the TSI pertains to wave vectors aligned with the flow, we set $Z_x=0$ in Eq. (\ref{eq:T}). The result is a tensor of the form,
\begin{equation}\label{eq:TS}
\mathfrak{T} = \left(\begin{array}{ccc}
      T_{11}   & T_{12} & 0 \\
      T_{12}^* & T_{22} & 0 \\
      0        & 0      & T_{33}
    \end{array}\right),
\end{equation}
where $z^*$ is the complex conjugate of $z$. The dispersion equation then reads,
\begin{equation}\label{eq:DisperTS}
T_{33}(T_{11}T_{22}-T_{12}^*T_{12})=0.
\end{equation}
The second factor can be further factorized, but the sub-factors still are 4th degree polynomials in $x$, with both even and odd powers of $x$. They are therefore difficult to deal with analytically. Yet, a numerical exploration shows that at least in the regime $B_0 \ll B_{cr}$, they do not yield unstable modes. The instability comes therefore from $T_{33}=0$, which is the usual dispersion equation for the TSI. In the present case with QED corrections, the equation derived in Ref. \cite{BretEPL2021} is recovered, namely,
\begin{equation}\label{eq:disperTSOK}
1+\xi-\frac{1}{\gamma_0^3 (x-Z_z)^2}-\frac{1}{\gamma_0^3 (x+Z_z)^2} = 0.
\end{equation}

We therefore here extend the results of Ref. \cite{BretEPL2021}: even when considering a full 3D system, the QED-corrected TSI is still 1D like.

\subsection{Filamentation instability (FI)}\label{sec:FIqed}
The FI pertains to wave vectors normal to the flow. We therefore set $Z_z=0$ in Eq. (\ref{eq:T}). A tensor of the form (\ref{eq:TS}) is obtained again, yielding a dispersion equation of the form (\ref{eq:DisperTS}).

In the classical case, the tensor elements $T_{33}$ is the one which yields the instability. Here it reads,
\begin{equation}\label{eq:T33Fila}
T_{33}=1+ \xi-\frac{1}{x^2}\left(\frac{2}{\gamma_0^3}+\frac{Z_x^2}{\beta_0^2}+\frac{1}{\gamma_0}\frac{2 Z_x^2}{ x^2-\Omega_B^2/\gamma_0^2}\right).
\end{equation}
In the case of the TSI, Eq. (\ref{eq:disperTSOK}) makes it possible to rescale  $x$ and $Z_z$ (or $\gamma_0$) and formally come back to the classical TSI dispersion equation \cite{BretEPL2021}. Such a procedure is not possible here. Still, some analytical conclusions can be reached.

For large $Z_x$, Eq. (\ref{eq:T33Fila}) gives the dispersion equation,
\begin{eqnarray}
  -\frac{1}{x^2}\left( \frac{Z_x^2}{\beta_0^2}\right)-\frac{2 \gamma_0 Z_x^2}{\gamma_0^2 x^4-x^2\Omega_B^2} &=& 0,  \nonumber  \\
  \Rightarrow   \frac{1}{\beta_0^2} + \frac{2 \gamma_0}{\gamma_0^2 x^2-\Omega_B^2} &=& 0 ,  \nonumber  \\
  \Rightarrow x^2 &=& \frac{\Omega_B^2-2 \beta_0^2 \gamma_0}{\gamma_0^2}.
\end{eqnarray}
This corresponds exactly to the classical growth rate (\ref{eq:FilaClass}). Therefore, QED effects do not affect the saturation value of the FI, nor the value of the magnetic field required to cancel it.

\begin{figure}
\begin{center}
 \includegraphics[width=.45\textwidth]{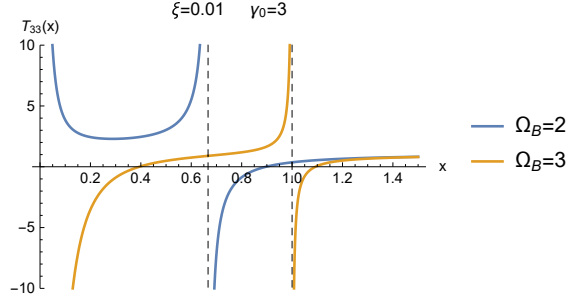}
\end{center}
\caption{Plot of $T_{33}(x)$ given by Eq. (\ref{eq:T33Fila}) for 2 values of $\Omega_B$. For the larger one, the equation $T_{33}(x)=0$ has 4 real roots whereas it has only 2 for the smaller $\Omega_B$.}\label{fig:t33}
\end{figure}

Regarding the stabilization of small $Z_x$'s, it can be directly derived from Eq. (\ref{eq:T33Fila}). Setting Eq. (\ref{eq:T33Fila}) to one single denominator will result in a fraction which numerator is a 4th degree polynomial in $x$. Therefore, $T_{33}=0$ must have 4 real roots for the system to be stable. We can then reason that,
\begin{itemize}
  \item $T_{33}(x)$ in an even function. So, if there are 2 positive real roots, there will also be 2 negative reals roots, hence a total of 4. We can therefore restrict the analysis to $x>0$.
  \item Then, $\lim_{x= + \infty} T_{33}(x)=1+\xi > 0$.
  \item Also, $\lim_{x=\Omega_B/\gamma_0^\pm} T_{33}(x)= \mp \infty$.
  \item Finally, $$\lim_{x=0^+} T_{33}(x) = -\mathrm{sign} \underbrace{\left(\frac{2}{\gamma_0^3}+\frac{Z_x^2}{\beta_0^2}-\frac{2 Z_x^2}{\Omega_B^2/\gamma_0}\right)}_{\equiv X}.$$
\end{itemize}
The situation is eventually summarized on Figure \ref{fig:t33}. When $\mathrm{sign}(X)<0$, $\lim_{x=0^+} T_{33}(x)=+\infty$ and the equation has but one positive real root. The system is therefore unstable. On the contrary, the system is stable with 2 positive real roots for the equation. The threshold pertains to $X=0$ which exactly gives back Eq. (\ref{eq:FilaStab}).

\begin{figure}
\begin{center}
 \includegraphics[width=.45\textwidth]{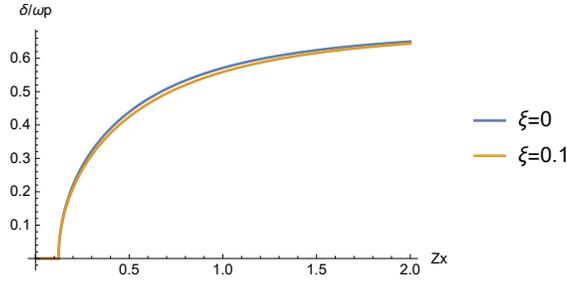}
\end{center}
\caption{QED corrected growth rate $\delta$ of the filamentation instability for $\Omega_B=1$ and two values of $\xi$.}\label{fig:filaqedOK}
\end{figure}

As previously said, the full dispersion equation for the FI is of the form (\ref{eq:DisperTS}). In the classical case, the factor $(T_{11}T_{22}-T_{12}^*T_{12})$ does not yield any instability. Here also, numerical exploration shows that as long as $\xi \lll 1$, this factor does not yield any unstable mode either.

We find therefore that QED effects do not affect the smallest unstable $Z_x$ of the FI, nor the saturation value of its growth rate for large $Z_x$'s.

Strictly speaking, the FI dispersion equation can be solved exactly since $T_{33}=0$, where $T_{33}$ is given by Eq. (\ref{eq:T33Fila}), yields a 4th degree polynomial with only even powers. Yet, we understand QED correction for any $Z_x$ cannot be considerable since the growth rate starts from a point independent of $\xi$, and ends up the same way. In this respect, Figure \ref{fig:filaqedOK} shows the growth rate of the FI for $\Omega_B=1$ and two values of $\xi$. A picture similar to the TSI one emerges: QED effects slightly reduce the growth rate.


\subsection{Oblique instabilities}
Surprisingly, what has been found for the TSI and the FI, namely that QED effects reduce the growth rate, is not systematically valid for oblique modes with both $Z_z \neq 0$ and $Z_x \neq 0$. Of course the 2 growth rates only slightly differ since $\xi \lll 1$, but the forthcoming analysis shows that in some regions of the $\mathbf{k}$ space, the QED corrected growth rate $\delta_{QED}$ can be \emph{larger} than the classical one $\delta_{Class}$.

We shall first prove analytically that $\delta_{Class} < \delta_{QED}$ for some oblique instabilities at large $k_\perp$ (i.e. $Z_x \gg 1$), before presenting a systematic numerical analysis of the difference.

\subsubsection{Oblique instabilities at $Z_x \gg 1$ and $Z_z=\Omega_B/\gamma_0$}
As evidenced on Fig. \ref{fig:class}, some oblique unstable modes are found for finite $Z_z$ and large $Z_x$. They were dubbed ``upper-hybrid-like'' modes in Ref. \cite{Godfrey1975}. Fig. \ref{fig:class} suggests that those found for the lower values of $Z_z$ vanish at large $Z_x$. This will be proved in the forthcoming analysis. Yet, those found at a larger $Z_z$ persist at $Z_x \gg 1$. They are centered around $Z_z = \Omega_B/\gamma_0$ \cite{Godfrey1975,BretPoPMagne}, which is a pole of the tensor $\mathfrak{T}_{Class}$ elements. We can see from Eqs. (\ref{eq:T},\ref{eq:Tqed}) that QED corrections do not modify this pole.

An analytical analysis of these unstable modes found at finite $k_\parallel$ (i.e. finite $Z_z$) and large $k_\perp$ ($Z_x \gg 1$) is possible with \emph{Mathematica}. One starts computing the determinant of the tensor $\mathfrak{T}$ given by Eq. (\ref{eq:Tqed}). A sum of several rational fractions follows. Setting them all on the same denominator results in a single fraction, which numerator is a polynomial $Q(x,Z_z,Z_x)$. The dispersion equation reads therefore $Q(x,Z_z,Z_x)=0$. $Q$ is a polynomial of degree 4 in $Z_x$. The dispersion equation for $Z_x \gg 1$ is therefore the coefficient $a_4$  of $Z_x^4$ in $Q$. It reads,
\begin{equation}\label{eq:a4}
a_4 = \gamma_0^3 (4 \xi-5) (x-Z_z)^2 (x+Z_z)^2~\left( \sum_{j=0}^4 b_j x^j \right),
\end{equation}
with,
\begin{widetext}
\begin{eqnarray}
  b_0 &=& \left(10 \gamma_0+(5-2 \xi) \Omega_B^2+\gamma_0^2 (2 \xi-5) Z_z^2\right)  \times       \left(2 \beta_0 ^2 \gamma_0-\Omega_B^2+\gamma_0^2 Z_z^2\right), \nonumber  \\
  b_1 &=& 0, \nonumber \\
  b_2 &=& 2 \gamma_0^2 \left[\gamma_0 \left(\beta_0 ^2 (2 \xi-5)+5\right)   +(5-2 \xi) \Omega_B^2+\gamma_0^2 (5-2 \xi) Z_z^2\right], \nonumber \\
  b_3 &=& 0, \nonumber \\
  b_4 &=& \gamma_0^4 (2 \xi-5) .
\end{eqnarray}

The first factor of $a_4$ does not yield any instability. As for the second one, the QED correction $\xi \lll 1$ enters its coefficients $b_j$ in such a way that they will only be slightly modified. And since the roots of a polynomial are continuous functions of its coefficients \cite{Uherka1977}, the associated growth rate is also only slightly modified.

Solving the equation allows to compute exactly the square of the growth rate. Taylor expanding it for $\xi \ll 1$ gives,
\begin{equation}
\delta = \left[\delta_{Class}^2 - \frac{2\xi}{5\gamma_0} \left(1-\frac{\beta_0^2+2 \gamma_0 Z_z^2+1}{\sqrt{\beta_0^4+\beta_0^2 \left(2-4 \gamma_0 Z_z^2\right)+4 Z_z^2 \left(\gamma_0+\Omega_B^2\right)+1}}\right) + \mathcal{O}(\xi^2) \right]^{1/2}.
\end{equation}
It is interesting to evaluate this expression for $Z_z=\Omega_B/\gamma_0$. The result is simply,
\begin{equation}
\delta(Z_z=\Omega_B/\gamma_0) = \left[\delta_{Class}^2 - \frac{2\xi}{5\gamma_0} \left(1-\underbrace{\frac{\beta_0^2+\frac{2 \Omega_B^2}{\gamma_0}+1}{\sqrt{-\frac{4 \left(\beta_0^2-1\right) \Omega_B^2}{\gamma_0}+\left(\beta_0 ^2+1\right)^2+\frac{4 \Omega_B^4}{\gamma_0^2}}}}_{\equiv  A/B}\right) + \mathcal{O}(\xi^2) \right]^{1/2}.
\end{equation}
\end{widetext}

Clearly, if $A/B > 1$, then the QED growth rate will be \emph{larger} than the classical one, instead of smaller as is the case for the TSI and the FI. Now, some algebra shows that,
\begin{eqnarray}
B^2-A^2 &=& -\frac{8 \beta_0^2 \Omega_B^2}{\gamma_0}  \nonumber \\
\Rightarrow \frac{A^2}{B^2} &=& 1 + \frac{1}{B^2}\frac{8  \beta_0^2 \Omega_B^2}{\gamma_0} > 1.
\end{eqnarray}

\begin{figure}
\begin{center}
\textbf{(a)}\\
 \includegraphics[width=.5\textwidth]{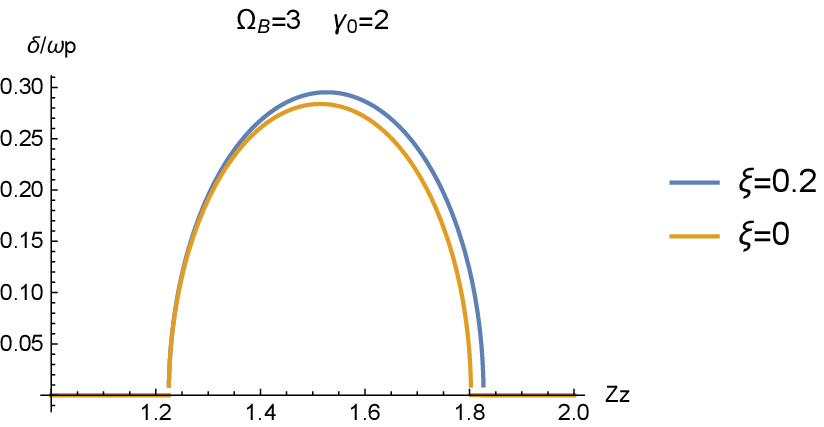}\\
 \textbf{(b)}\\
 \includegraphics[width=.5\textwidth]{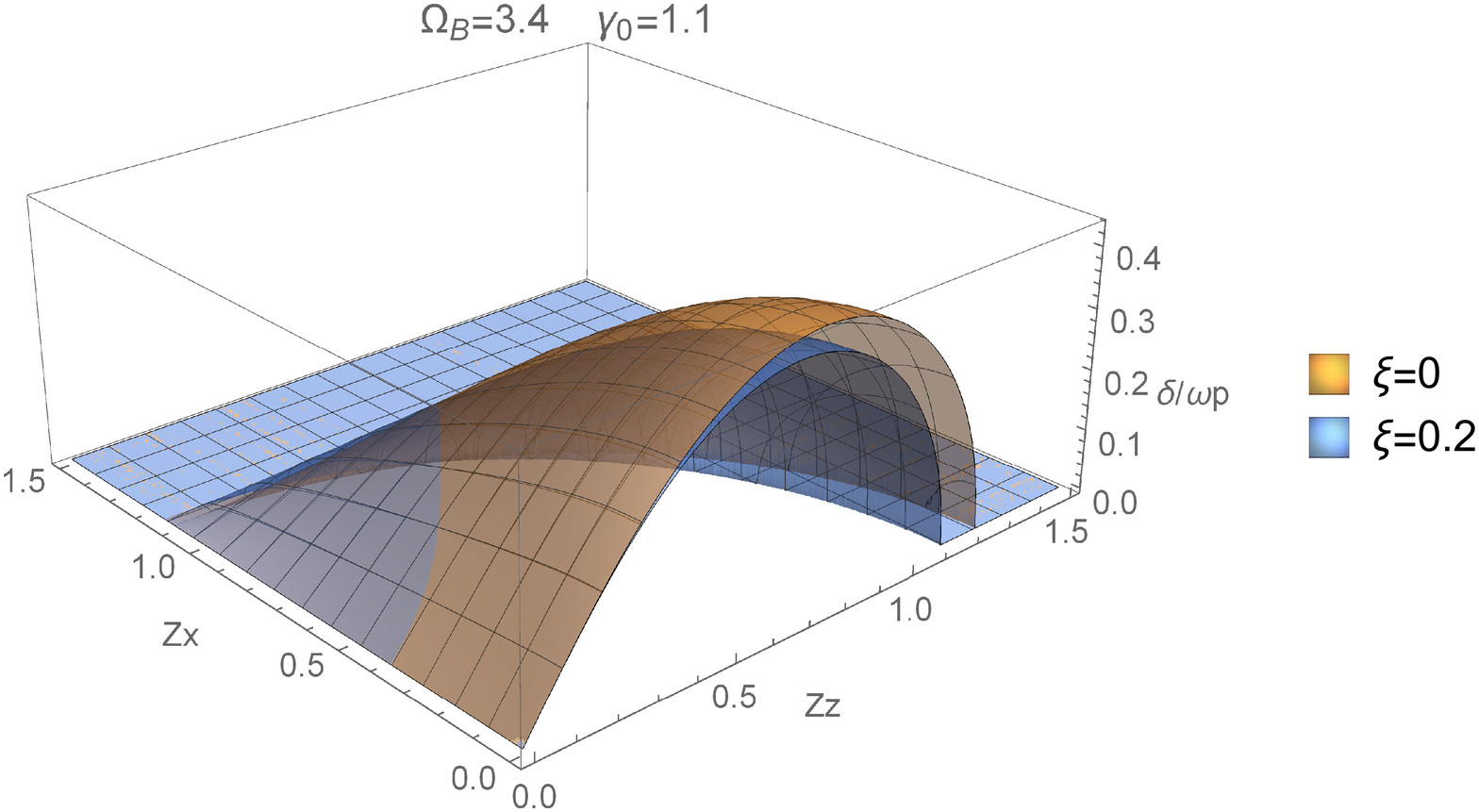}
\end{center}
\caption{\textbf{(a)}: Growth rate of the oblique modes around $Z_z=\Omega_B/\gamma_0$ and $Z_x=\infty$. The QED corrected growth rate is larger than the classical one. \textbf{(b)}: Typical growth rate for the QED ($\xi=0.2$) and the classical cases when the FI is stabilized with $\Omega_B > \Omega_{Bc}$. The value of $\xi$ is exaggerated in both plots to make QED effects visible.}\label{fig:GdZx}
\end{figure}

 Therefore, for $Z_z=\Omega_B/\gamma_0$ and $Z_x\rightarrow\infty$, the QED corrected growth rate is larger than the classical one. This is illustrated on Figure \ref{fig:GdZx}-(a) which shows the two quantities as a function of $Z_z$. Note that temperature effects are likely to stabilize these large $k_\perp$ modes \cite{Silva2002,BretPoPReview}. Note also that since Eq. (\ref{eq:a4}), that is, $a_4=0$,  features only 1 unstable mode, it means that out of the 2 branches visible at large $Z_x$ on Fig. \ref{fig:class}, only one persists in the limit $Z_x=\infty$. We just found that this is the one located around $Z_z =\Omega_B/\gamma_0$.

We now show numerically that $\delta_{Class} < \delta_{QED}$ does not occur only at $Z_z=\Omega_B/\gamma_0$ and $Z_x\rightarrow\infty$.

\begin{widetext}
\begin{figure}
\begin{center}
\includegraphics[width=\textwidth]{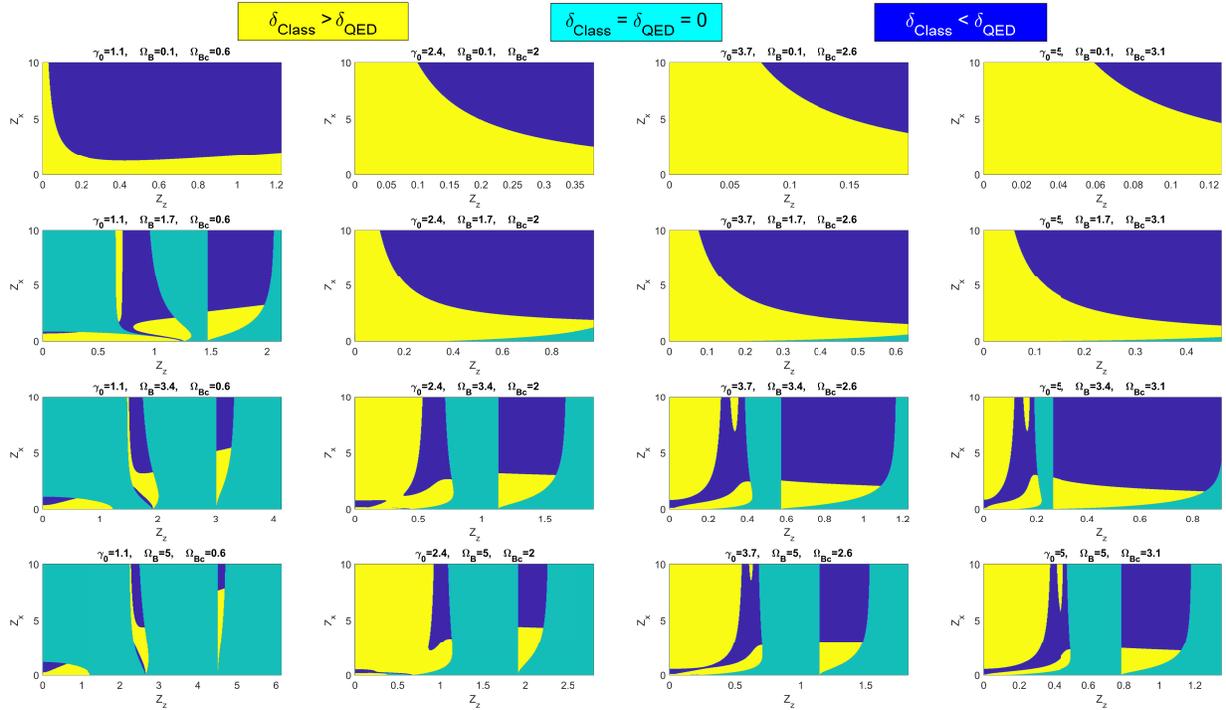}
\end{center}
\caption{Plot of $\delta_{Class} - \delta_{QED}$ for $\xi=10^{-2}$ and various combinations $(\gamma_0,\Omega_B)$. The quantity $\Omega_{Bc}$ indicates the threshold (\ref{eq:OmegaBc}) for the stabilization of the FI. While the classical growth rate is larger than the QED one for the TSI and the FI, the contrary can happen for oblique modes (dark blue regions).}\label{fig:maps}
\end{figure}
\end{widetext}

\subsubsection{Numerical study}
We now systematically study the difference $\delta_{Class} - \delta_{QED}$ over the whole $\mathbf{k}$ space, for several values of $\gamma_0$ and $\Omega_B$. The result is displayed on Figure \ref{fig:maps}. Note that for faster computation, the polynomial dispersion equation has been transferred from \emph{Mathematica} to \emph{MatLab} using the procedure described in Ref. \cite{BretMatLab2010}. For each combination $(\gamma_0,\Omega_B)$, the value of the critical magnetic parameter $\Omega_{Bc}$ [Eq. (\ref{eq:OmegaBc})] cancelling the filamentation instability is indicated on Fig. \ref{fig:maps}.

These calculations confirm what was previously found. Not only $\delta_{Class} < \delta_{QED}$ is fulfilled around $Z_z=\Omega_B/\gamma_0$ and at large $Z_x$'s, but also in several parts of the spectrum. In some cases, the dark blue regions, which indicates places where $\delta_{Class} < \delta_{QED}$, seems to reach the vertical axis $Z_z=0$. One could then think that in such cases we have $\delta_{Class} < \delta_{QED}$ for the FI, in contradiction with the conclusions of Section \ref{sec:FIqed}.

Yet, as shown for example on Fig. \ref{fig:GdZx}-(b), this is not the case. On this plot, the FI is stabilized since $\Omega_B=3.4 > \Omega_{Bc}=0.6$. While we retrieve $\delta_{Class} > \delta_{QED}$ for the TSI, there is a region near $0.5 \lesssim Z_x \lesssim 1$ where $\delta_{Class} < \delta_{QED}$, while both go to 0 simultaneously for $Z_z=0$ since the criteria for canceling them is the same. In other words, we do not find here $\delta_{Class}(Z_x=0) < \delta_{QED}(Z_x=0)$. Instead, we find cases with $\delta_{Class}(Z_x=0^+) > \delta_{QED}(Z_x=0^+)$, and $\delta_{Class}(Z_x=0) = \delta_{QED}(Z_x=0) =0$.

\section{Conclusion}
We computed the QED corrections to counter-streaming instabilities resulting from harmonic perturbations with any possible orientation of the wave vector.

As was the case for the TSI, finite first order corrections demand the presence of a flow-aligned static magnetic field $\mathbf{B}_0$. If $B_0=0$, the fields cannot reach high enough intensities when growing from 0 during the linear phase, for QED effects to appear.

Note that QED effects could arise during the \emph{non}-linear phase of an \emph{un}-magnetized system if the field at saturation approaches the Schwinger limit $B_{cr}$. For example, the saturation value $B_{sat}$ of the magnetic field for the filamentation instability is of the order of \cite{davidsonPIC1972,BretPoP2013},
\begin{equation}
B_{sat} = \sqrt{\gamma_0}\frac{mc\omega_p}{q},
\end{equation}
so that $B_{sat} \sim B_{cr}$ gives,
\begin{eqnarray}
\sqrt{\gamma_0}\frac{mc\omega_p}{q} \sim \frac{m^2c^3}{q\hbar } ~~&\Rightarrow& ~~\gamma_0 \sim  \frac{1}{\omega_p^2} \frac{m^2c^4}{\hbar^2},  \nonumber \\
&\Rightarrow& ~~\gamma_0 \sim \left( \frac{\ell_e}{\ell_C} \right)^2,
\end{eqnarray}
where $\ell_e=c/\omega_p$ is the electron inertial length and $\ell_C=\hbar/mc$ the reduced electronic Compton wavelength. With $\ell_C=3.8\times 10^{-11}$ cm and $\ell_e=5.3 \times 10^5 n_e^{-1/2}$ cm, where $n_e$ is the electronic density in $\mathrm{cm}^{-3}$ \cite{NRL2019}, this translates to,
\begin{equation}
\gamma_0 \sim \frac{1.9\times 10^{32}}{n_e ~ [\mathrm{cm}^{-3}] }.
\end{equation}

Hence, $B_{sat} \sim B_{cr}$ could be achieved for extremely high Lorentz factors and/or beams densities.

Previous results for the TSI are recovered. In addition, we find here that even in 3D, the QED corrected TSI remains a 1D problem.

Choosing the wave vector perpendicular to the flow allows to analyze the FI. We find that QED effects neither change the smallest unstable $k_\perp$ nor the growth rate at large $k_\perp$. In between, QED corrections slightly decrease the growth rate.

Noteworthily, when it comes to oblique unstable modes, analytical and  numerical calculations do \emph{not} confirm the trends found for the TSI and the FI. While the growth rate reductions can be attributed to virtual particles screening the charges (see \cite{weinberg2005quantum}, p. 482), its increase in some regions of the $\mathbf{k}$ space is surprising.

Some important effects have been left out in this article. When particles oscillate in the first order growing fields, they may emit gamma photons which in turn trigger pair production \cite{GrismayerPoP2016,GrismayerPRE2017}. These effects should arguably be worked out in future works.

\section{Acknowledgments}
A.B. acknowledges support by grants ENE2016-75703-R from the Spanish
Ministerio de Ciencia, Innovaci\'{o}n y Universidades and SBPLY/17/180501/000264 from the Junta de
Comunidades de Castilla-La Mancha. Thanks are due to Luis Silva and Thomas Grismayer for valuable
inputs.


\end{document}